\documentclass[sigconf]{aamas}  % do not change this line!

\usepackage{times}
\usepackage{soul}
\usepackage{url}
\usepackage[utf8]{inputenc}
\usepackage{graphicx}
\usepackage{amsmath}
\usepackage{amsthm}
\usepackage{algorithm}
\usepackage{algorithmic}
\usepackage{multirow}
\usepackage{bm}
\usepackage{booktabs}

\usepackage{flushend}
\setcopyright{ifaamas}  % do not change this line!
\acmDOI{doi}  % do not change this line!
\acmISBN{}  % do not change this line!
\acmConference[AAMAS'20]{Proc.\@ of the 19th International Conference on Autonomous Agents and Multiagent Systems (AAMAS 2020), B.~An, N.~Yorke-Smith, A.~El~Fallah~Seghrouchni, G.~Sukthankar (eds.)}{May 2020}{Auckland, New Zealand}  % do not change this line!
\acmYear{2020}  % do not change this line!
\copyrightyear{2020}  % do not change this line!
\acmPrice{}  % do not change this line!

\begin{document}

\title{Directionality Reinforcement Learning \\to Operate Multi-Agent System without Communication}  % put your title here!
%\titlenote{Produces the permission block, and copyright information}

% AAMAS: as appropriate, uncomment one subtitle line; check the CFP
%\subtitle{Extended Abstract}
%\subtitle{Blue Sky Ideas Track}
%\subtitle{JAAMAS Track}
%\subtitle{Demonstration}
%\subtitle{Doctoral Consortium}

% AAMAS: submissions are anonymous for most tracks
%\author{Paper \#XXX}  % put your paper number here!

\author{Fumito Uwano}
\orcid{0000-0003-4139-2605}
\affiliation{%
  \institution{Okayama University}
  \streetaddress{3-1-1, Tsushima-naka, Kita-ku}
  \city{Okayama} 
  \postcode{700-8530}
}
\email{uwano@cs.okayama-u.ac.jp}

\author{Keiki Takadama}
\affiliation{%
  \institution{The University of Electro-Communications}
  \streetaddress{1-5-1, Chofugaoka, Chofu-shi}
  \city{Tokyo} 
  \postcode{182-8585}
}
\email{keiki@inf.uec.ac.jp}

\begin{abstract}  % put your abstract here!
This paper establishes directionality reinforcement learning (DRL) technique to propose the complete decentralized multi-agent reinforcement learning method which can achieve cooperation based on each agent's learning: no communication and no observation. Concretely, DRL adds the direction ``agents have to learn to reach the farthest goal among reachable ones" to learning agents to operate the agents cooperatively. Furthermore, to investigate the effectiveness of the DRL, this paper compare Q-learning agent with DRL with previous learning agent in maze problems. Experimental results derive that (1) DRL performs better than the previous method in terms of the spending time, (2) the direction makes agents learn yielding action for others, and (3) DRL suggests achieving multi-agent learning with few costs for any number of agents.
\end{abstract}

\keywords{Multi-Agent System; Reinforcement Learning; Internal Reward; Intrinsic Motivation}  % put your semicolon-separated keywords here!

\maketitle

%%%%%%%%%%%%%%%%%%%%%%%%%%%%%%%%%%%%%%%%%%%%%%%%%%%%%%%%%%%%%%%%%%%%%%%%%%%%%%%%%%%%%%%%%%%%%%%%%%%%%%%%%
%% start of main body of paper

\section{Introduction}
Multi-agent reinforcement learning (MARL) is the important study in this human society: A lot of agents solve the problem with cooperation together, {\it e.g.}, robot logistic system, traffic light control and more \cite{GodoyAAAI2016,LiuITSC2017}. Generally, the most important factor is communication with agents for cooperation. Lately, other agent modeling based on the communicated information becomes important, {\it e.g.}, \cite{AlbrechtAIJ2018}. These techniques make an agent model other agent, predict it's action and purpose, and act by the cooperative policy based on the action and the purpose. Although these techniques achieve the agents' complex cooperation, a lot of information is required. This suggests that the cooperation cannot be achieved in some situations, that is, non- or less-communication like field robotics or a lot of agents. In addition, non-communicative cooperation by reinforcement learning is challenging and important topic.
For example, Raileanu et al. proposed cooperative action learning method by modeling the opponent agent (Self-other modeling: SOM). Zemzem and Tagina's method (Cooperative multi-agent learning approach based on the most recently modified transitions: CMRL-MRMT) make agents learn cooperative action by copying the learning result of the advantageous agent \cite{RaileanuICML2018,ZemzemMDAI2015}. These methods utilize the information like the agents' states, actions, learning results, and more to achieve the cooperation. However, both methods cannot cooperate without communication.
\if 0
\begin{table}[t]
\begin{center}
\caption{Addressing of previous methods \label{tab:addressing}}
\begin{tabular}{|c||c|c|} \hline
&\multicolumn{2}{c|}{Communication} \\
&On&Off \\ \hline
\multirow{3}{*}{Advising}&\cite{RaileanuICML2018}& \\
&\cite{LittmanICML1994}&\cite{UwanoJCMSI2018} \\
&\cite{ShiraishiJCMSI2019}& \\ \hline
\multirow{3}{*}{Co-operation}&\cite{ZemzemMDAI2015}& \\
&\cite{AraiICSAI2000}&\cite{UwanoJCMSI2019} \\
&\cite{MiyazakiJACIII2017}&\\ \hline
\end{tabular}
\end{center}
\end{table}
\fi

On the other hand, some researchers aim to propose cooperative learning methods without communication. However, these methods are not complete, that is, they have to be under some conditions. For example, though Arai and Katia mentioned the non-communicative method, the agent can observe other agent's state \cite{AraiICSAI2000}, and Shiraishi et al. extended that method, but the condition is remain required \cite{ShiraishiJCMSI2019}. Uwano et al. adds some direction to learning agents to cooperate each other without communication (Profit minimizing reinforcement learning: PMRL) \cite{UwanoJCMSI2018}, but PMRL cannot make the cooperation in any cases. Thus, a complete non-communicative reinforcement learning method for multi-agent cooperation cannot be proposed.
To tackle this problem, we aim to propose the complete decentralized multi-agent learning method. 
We focus on the direction in PMRL is important. That is because the other methods of PMRL translate the communication to the observation, and the direction is only way to propose the non-communicative method. For example, since we already have the rule ``To reach a goal in a maze, you have to go with your right side on the wall", we can reach a goal of new maze. This rule is the direction in PMRL. Concretely, we extend PMRL and establish a directionality reinforcement learning as the knowledge reusing technique without knowledge.

The remainder of this paper is organized in the following manner. The section \ref{sec:mdp} explains MDP in this paper
%, and the section \ref{sec:background} introduces the background techniques to establish the directionality reinforcement learning
. The section \ref{sec:pmrl} explains the mechanisms of PMRL. After that, the section \ref{sec:drl} proposes the directionality reinforcement learning. The section \ref{sec:experiment} discusses the experiment and the results. Finally, our conclusions are summarized in Section \ref{sec:conclusion}.
\section{Simulated markov decision process} \label{sec:mdp}
Reinforcement learning performs on Markov decision process (MDP). This paper simulates the general MDP, but some parts of the simulated MDP are different. The equations (\ref{eq:A}), (\ref{eq:tau}) and (\ref{eq:r_iom}) show the rules of the MDP. $S$ is a state, $A$ is a joint action as a direct product of all actions of all agents calculated by the equation (\ref{eq:A}). The MDP transfers itself to the next state with the probability being 1 by the equation (\ref{eq:tau}). During the cycle of that transfer, the MDP gives each agent the reward calculated by the equation (\ref{eq:r_iom}). $r$ is the reward value, $s_i$ is the state of the agent $i$, and $S_{goal}$ is the set of the goal states. The equation (\ref{eq:r_iom}) indicates if the agent could reach the goal at first, it can acquire the reward.
\begin{flalign}
&A = a_1 \times a_2 \times ... \times a_n \label{eq:A}\\
&\tau: S \times A \times S \rightarrow 1 \label{eq:tau}\\
&R^{om}_i = \left\{ r  | s_i, s_{-i} \in S_{goal} \land s_i \neq s_{-i} \right\} \label{eq:r_iom}
%\{r_{s^i} | s^i \in S_{goal}\} \label{eq:r^i}
\end{flalign}
\if 0
\section{Background} \label{sec:background}
Silva et al. introduced the knowledge reusing techniques for multi-agent cooperation: Knowledge from previous tasks, Learning from Demonstration, Imitation Learning, Advising, Curriculum Learning and Additional Related Areas \cite{SilvaIJCAI2018}. 
To achieve the non-communicative learning for multi-agent system, the curriculum learning is required.
\subsection{Curriculum learning}

However, the curriculum learning method cannot be optimal for cooperation with the least information.
\subsection{Reward shaping}
Uwano proposed the non-communicative multi-agent cooperative learning method (Profit minimizing reinforcement learning: PMRL) \cite{UwanoJCMSI2018}, but PMRL requires the little communication in some situations.
\subsection{Directionality learning}
\fi
\section{Profit minimizing reinforcement learning (PMRL)} \label{sec:pmrl}
PMRL forces the cooperation by setting the internal reward. Concretely, PMRL selects an appropriate goal for each agent and makes the agent learn to reach the goal. To realize that, PMRL employed an internal reward and a goal value. The internal reward is the reward inside the agent calculated from an external reward. The goal value is the selection worth of the goal to reach it with cooperation.
The agent updates Q-values based on the internal rewards in each step, and updates the goal values and the internal rewards in each episode. The mechanisms of the goal value setting and the internal reward setting are explained as the following subsections.
\subsection{Goal value}
The goal value is calculated from the minimum number of the steps to reach each goal. The equation (\ref{eq:bidpmrl}) is the update function of the goal value. If the agent reach the goal $g$ the earliest, the goal value is updated by the upper function; otherwise, it is updated by the lower function. $bid_g$ is the goal value of the arbitrary goal $g$. $n_g$ is an update number of the goal value $bid_g$. The agent calculates the averagely value of the minimum number of steps from the initial iteration to the current one. Since $t_g$ is added into the update function while only the situation where the agent has reached the goal $g$ earliest, $bid_g$ is the minimum number of steps to reach the goal $g$. This suggests that the agent learns to reach the farthest goal.
\begin{align}
bid_g \leftarrow \left\{ \begin{array}{cc}
\frac{n_g - 1}{n_g} bid_g + \frac{t_g}{n_g} &if\ first\ reaching \label{eq:bidpmrl} \\
\frac{n_g - 1}{n_g} bid_g + \frac{0}{n_g} &otherwize
	\end{array}
	\right.
\end{align}
\subsection{Internal reward and learning}
%Q学習の説明でQ値の収束値を書く
The internal reward is calculated by the learning agent to reach the goal $g$ using the equation (\ref{eq:irpmrl}). $ir(g)$ is the internal reward, $\gamma$ is the discount rate ($0<\gamma<1$), and the goal $g_o$ is the other goal.
The agent estimates the Q-value using Eq.(\ref{eq:qvpmrl}). $s$ and $a$ are an arbitrary state and an arbitrary action, and $s^{\prime}$ and $a^\prime$ are the next state and the next action. $\alpha$, $\gamma$ and $\delta$ are the learning rate ($0<\alpha<1$), the discount rate and the positive constant value. Note that if $\delta$ do not exist, the agent learns to reach the goal $g_o$.
\begin{align}
	&ir(g) = \left\{ \begin{array}{cc}
	\underset{{g_{o} \in G, g_{o} \neq g}}{\max} r_{g_{o}}\gamma^{t_{g_{o}}-t_{g}}+\delta & if\ first\ reaching \\
	0 & otherwise
	\end{array}
	\right.
	\label{eq:irpmrl} \\
	&Q(s,a) \leftarrow \nonumber \\
	&\ \ \ \ (1-\alpha)Q(s,a)+\alpha\left[ ir(g)+\gamma\max_{a^\prime \in A} Q(s^\prime,a^\prime) \right] \label{eq:qvpmrl}
\end{align}
\section{Directionality Reinforcement Learning (DRL)} \label{sec:drl}
To achieve non-communicative multi-agent cooperation, we add the direction to agents' learning, which is named ``Directionality Reinforcement Learning (DRL)". PMRL can direct learning agents to reach the farthest goal, but Uwano did not mention any ways to achieve the condition ``if first reaching or not" in PMRL's mechanisms. DRL solve this problem to establish the directionality technique. Concretely, DRL utilizes the mean value of the acquired rewards.
\subsection{Mechanism}
The added direction is that agents have to learn to reach the farthest goal among reachable ones. The equation (\ref{eq:condition}) calculates a condition value to determine the first reaching or not (reachable goal or not). Concretely, the equation calculates the mean value of the acquired rewards through whole episodes. Since the agent getting the reward has reached the goal at first, a lot of mean value indicates the goal indicates the first reaching (reachable goal).
\begin{align}
R_{g,ave} = \frac{1}{n_g} \sum_e^{n_g} R^{om}_i(e) \label{eq:condition}
\end{align}
This mechanism with the goal value in PMRL can be the appropriate direcionality. Hence, DRL combines the above mechanism with the goal value calculation as follows: 
\begin{align}
bid_g \leftarrow \left\{ \begin{array}{cc}
\frac{n_g - 1}{n_g} bid_g + \frac{t_g}{n_g} &if\ R_{g,ave} > Th \label{eq:biddrl} \\
\frac{n_g - 1}{n_g} bid_g + \frac{0}{n_g} &otherwize
	\end{array}
	\right.
\end{align}
Where $Th$ is a threshold to determine the first reaching. DRL determines the first reaching by the equation (\ref{eq:condition}).
\subsection{Algorithm}
Algorithm \ref{alg:pmrl} is a whole algorithm of DRL.
Q-value is initialized at first (line 1). $\bm{t}, \bm{bid}, \bm{R_{ave}}$ and $\bm{c}$ are arrays of minimum number of steps, goal values, averaged rewards and those update counts, respectively (line 2). Before the learning, the state is initialized as $s_0$, the goal $g_{sel}$ as the agent has to reach by the internal reward is set randomly, and the goal set $S_{end}$ is set (lines 3, 4, 5).
After that, the agent observes their own state, selects and executes an action, receives a reward, and observes a next state $s^\prime$ (lines from 6 to 8). After the observation, the agent calculates the internal reward $ir(g_{sel})$ to reach the goal $g_{sel}$, and updates the Q-value from the internal reward (lines 9 and 10). If the observed state $s^\prime$ is included in the goal set $S_{end}$, the agent breaks the loop storing the acquired reward to $\bm{R_{ave}}$ and updating the update count $\bm{c}$ (lines from 11 to 15). After the steps, if the agent has reached the goal $s^\prime$ and the current $step$ is the smaller than before, it adds $step$ to the array $\bm{t}$ (liens 17, 18 and 19). At last, the agent select the goal $g_{sel}$ and updates the goal value $\bm{bid}$ (lines from 20 to 29). The agent selects the goal $g_{sel}$ with the max value among the arrays $\bm{bid}$, and $g_{sel}$ can be selected randomly with a certain probability. Furthermore, the agent updates the goal value how is updated by the top of the equation (\ref{eq:biddrl}) only when the reaching the goal $g_{sel}$ at first with the value $\frac{\bm{R_{ave}}[g_{sel}]}{\bm{c}[g_{sel}]}$ being over the threshold $Th$.
\begin{algorithm}[t] 
\caption{DRL}         
\label{alg:pmrl}
\algsetup{
linenosize=\small,
linenodelimiter=.
}                          
\begin{algorithmic}[1]                  
\STATE $Q(s,a)$ is initialized, $\forall s \in S, \forall a \in A$
\STATE Array of minimum number of steps $\bm{t}$ and array of goal values $\bm{bid}$, array of averaged rewards $\bm{R_{ave}}$, and array of update counts $\bm{c}$
\STATE Initial state $s_0$, the goal $g_{sel}$ is initialized, and array of goals ${S_{end}}$
\FOR{$iteration = 1$ to $MaxIteration$}
\STATE $s = s_0$
\FOR{$step = 1$ to $MaxStep$}
\STATE $a = ActionSelect(Q, s)$
\STATE Executing the action $a$, acquiring the reward $R$, and observing the next state $s^\prime$
\STATE Calculating the internal reward $ir$ to reach the goal $g_{sel}$ from the reward
\STATE {$Q(s,a) = (1-\alpha)Q(s,a) + \alpha\left[ir(g_{sel})+\gamma\underset{a^\prime \in A}{\max}Q(s^\prime,a^\prime)\right]$}
\IF {$s^\prime=state_g \in S_{end}$}
\STATE {$\bm{R_{ave}}[g] = \bm{R_{ave}}[g]+R$}
\STATE {$\bm{c}[g]++$}
\STATE break
\ENDIF
\ENDFOR
\IF{$s^\prime \in s_{end} \land step < \bm{t}[g]$}
\STATE $\bm{t}[g] = step$
\ENDIF
\IF {$\bm{bid}[g]$ is the largest}
\STATE $g_{sel} = g$
\ENDIF
\STATE Determining $g_{sel}$ randomly with a certain probability
\STATE $n^i_{g_{sel}}++$
\IF {first reaching the goal $g_{sel}$ and $\frac{\bm{R_{ave}}[g_{sel}]}{\bm{c}[g_{sel}]}>Th$}
\STATE {$\bm{bid}[g_{sel}]=\frac{n_{g_{sel}} - 1}{n_{g_{sel}}}\bm{bid}[g_{sel}]+\frac{\bm{t}[g_{sel}]}{n_{g_{sel}}}$}
\ELSE
\STATE {$\bm{bid}[g_{sel}]=\frac{n_{g_{sel}} - 1}{n_{g_{sel}}}\bm{bid}[g_{sel}]+\frac{0}{n_{g_{sel}}}$}
\ENDIF
\ENDFOR
\end{algorithmic}
\end{algorithm}
\section{Experiment} \label{sec:experiment}
\subsection{Experimental design \& setting}
To investigate the effectiveness of the improved DRL, we compares the performance of DRL and Profit sharing (PS) for multi-agent system in maze problems explained after. The employed problems are 20 kinds with randomly initialized start and goal locations for two
%, three
 and five agents. This paper evaluates the number of steps until all agents have reached the goals, that is, the smaller number of steps are better. Note that this paper evaluates that as the maximum number of steps when the agents have reached the same goal together. 

The experiment tried with 10 seeds for random selection. As for the parameter, the learning iterations are limited to 50000 for two agents, 
%100000 for three ones 
and 500000 for five ones. The steps are limited to 100, the initialized Q-value is 0, the learning rate $\alpha$ and the discounted rate $\gamma$ of Q-learning is 0.1 and 0.9, respectively. The external reward is 10. The constant value $\delta$ is 10, and the threshold $Th$ is 5.
\if 0
\begin{table}[b]
\begin{center}
\caption{パラメータ \label{tab:pmrlomexperiment2parameter}}
\begin{tabular}{|c|c|c|c|} \hline
&\multicolumn{3}{c|}{DRL} \\ \hline
  &two agents&three agents&five agents \\ \hline
	エピソード数&50000&100000&500000 \\ \hline
最大ステップ数&\multicolumn{3}{c|}{100} \\ \hline
初期Q値&\multicolumn{3}{c|}{0} \\ \hline
学習率$\alpha$&\multicolumn{3}{c|}{0.1} \\ \hline
割引率$\gamma$&\multicolumn{3}{c|}{0.9} \\ \hline
%random seed&\multicolumn{6}{c|}{0 ﾂ～ 29} \\ \hline
報酬値&\multicolumn{3}{c|}{10} \\ \hline
	定数 $\delta$&\multicolumn{3}{c|}{10} \\ \hline
	閾値 $Th$&\multicolumn{3}{c|}{5} \\ \hline
\end{tabular}
\end{center}
\end{table}
\fi
PS has the different reward function against DRL and that function indicates the equation (\ref{eq:r_ips}). $r$ is the reward value and $N$ is the episode number to reach the goal. PS gives the reward value $\frac{r}{N}$ to all agents at the end of the current iteration. Since PS utilizes the global information $N$, {\it i.e.}, agents know an appropriate situation for cooperation, PS is advantageous than DRL.
\begin{flalign}
&R^{ps}_i = \left\{\frac{r}{N} | s_i, s_{-i} \in S_{goal} \land s_i \neq s_{-i} \right\} \label{eq:r_ips}
%\{r_{s^i} | s^i \in S_{goal}\} \label{eq:r^i}
\end{flalign}
\subsection{Maze problem}
\begin{figure}[tb]
\begin{center}
\includegraphics[width=0.45\textwidth]{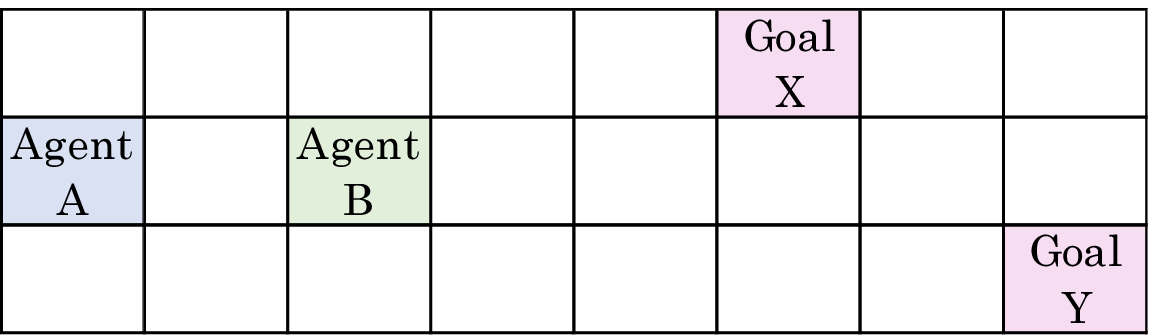}
\caption{A maze problem for two agents \label{fig:multiagentmaze}}
\end{center}
\end{figure}

This paper employs maze problems like the grid world in the figure \ref{fig:multiagentmaze}. The agent can move up, down, left, and right, and several agents can observe the same state together. In addition, each agent can reach both goals, but if the agent has reached the goal, it repeats staying this goal and cannot move from there. The agent can acquire the rewards if it reaches the goal at first, and if the several agents have reached the same goal, some one reached that at first cannot acquire the reward.
Generally, if two agents learn to reach the goals with Q-learning and Profit sharing on the grid world shown in Figure \ref{fig:multiagentmaze}, both agents learn to reach the goal X together and make a conflict. In this situation, the agents A and B should learn to reach the goals X and Y, respectively. Therefore, we can check whether the agents can behave cooperatively or not by determining which this situation arises or not.
\subsection{Results}
The table \ref{tab:mediansteps} shows minimum values of spent steps in all mazes. The top indicates the results for two agents, which the bottom indicates those for five agents, and the top and the bottom in each case indicate the results of DRL and PS, respectively. Each column indicates each case, and each cell includes the step which bold type indicates the better performance than the opponent. From these results, DRL performs better than PS in all maze problems. Since PS can give all agents the maximum value of rewards, that is, PS tells them the best situation for cooperation unlike DRL, this fact is very important. DRL can enable agents to learn cooperative actions without communication. 
\begin{table*}[tb]
\begin{center}
\caption{Minimum values of reached steps at the end of learning \label{tab:mediansteps}}
\begin{tabular}{|c|c||c|c|c|c|c|c|c|c|c|c|} \hline
&&maze 1&maze 2&maze 3&maze 4&maze 5&maze 6&maze 7&maze 8&maze 9&maze 10 \\ \hline
\multirow{2}{*}{Two agents}&DRL&\textbf{7}&\textbf{5}&\textbf{3}&\textbf{6}&\textbf{5}&\textbf{3}&\textbf{4}&\textbf{7}&\textbf{6}&\textbf{7} \\ \cline{2-12}
&PS&\textbf{7}&\textbf{5}&\textbf{3}&7&\textbf{5}&\textbf{3}&\textbf{4}&100&\textbf{6}&\textbf{7} \\ \hline \hline
%\multirow{2}{*}{Three agents}&DRL&5&3&6&55.5&4&2&4&3&5&6 \\ \cline{2-12}
%&PS& &&&&&&&&& \\ \hline \hline
\multirow{2}{*}{Five agents}&DRL&\textbf{9}&\textbf{9}&\textbf{15}&\textbf{7}&\textbf{13}&\textbf{15}&\textbf{9}&\textbf{7}&\textbf{10}&\textbf{16} \\ \cline{2-12}
&PS&100&100&100&100&100&100&100&100&100&100 \\ \hline
\end{tabular}
\end{center}
\end{table*}
\subsection{Discussion}
\begin{figure*}[tb]
\begin{center}
\includegraphics[width=0.9\textwidth]{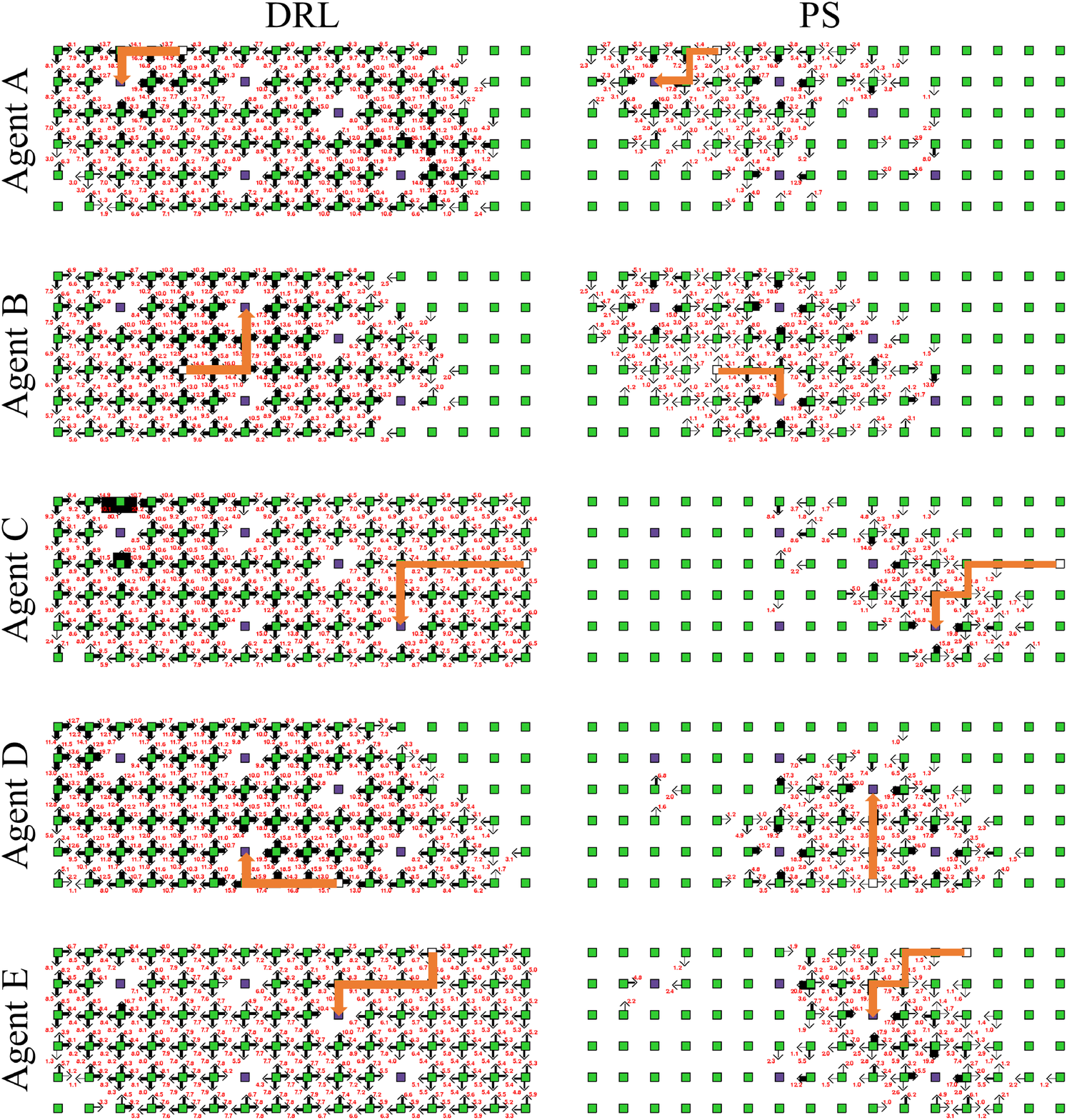}
\caption{Q-tables in the maze 4 for five agents \label{fig:qtables45}}
\end{center}
\end{figure*}

We discuss some characteristic result in the table \ref{tab:mediansteps}.
\if 0
\subsubsection{The maze 5 for two agents}
\begin{figure*}[tb]
\begin{center}
\includegraphics[width=0.9\textwidth]{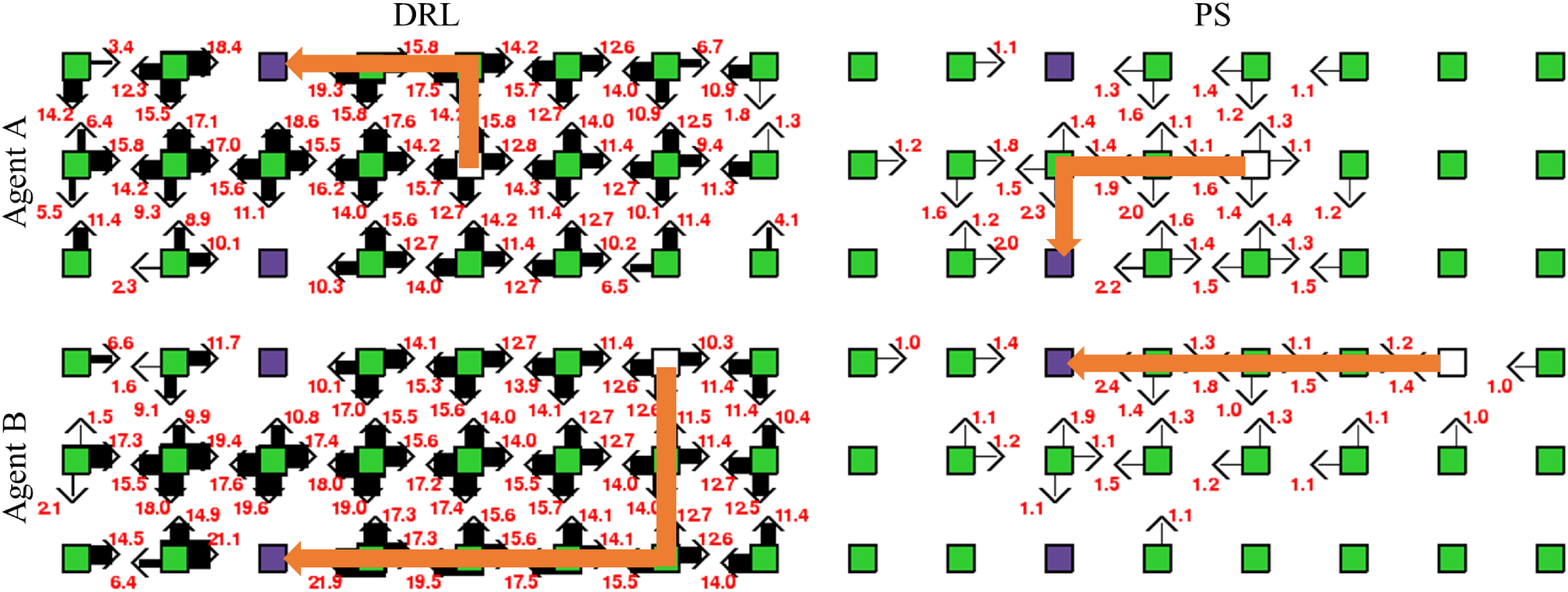}
\caption{Q-tables in the maze 5 for two agents \label{fig:qtables52}}
\end{center}
\end{figure*}
The figure \ref{fig:qtables52} shows Q-tables for all agents in the maze 5 for two agents. The left and the right sides indicate DRL and PS, while the top and the bottom indicate the agents A and B, respectively. The white and the violet squares shows the locations of the starts and the goals, respectively. The red values are the Q-values and the black arrows nearby those indicate the actions. The orange arrows are the trajectories from the start to the goal when the agent keeps selecting the action with the maximum Q-value. From this figure, DRL adds the different direction to the agents from PS, and that directionality causes the spent time of DRL is more than that of PS.
\subsubsection{The maze 4 for five agents}
\fi
The figure \ref{fig:qtables45} shows Q-tables for all agents in the maze 4 for five agents. The left and the right sides indicate DRL and PS, while the top and the bottom indicate the agents A and B, respectively. The white and the violet squares shows the locations of the starts and the goals, respectively. The red values are the Q-values and the black arrows nearby those indicate the actions. The orange arrows are the trajectories from the start to the goal when the agent keeps selecting the action with the maximum Q-value. From this figure, DRL makes all agents learn to avoid a conflict, on the other hand, PS could not make that. Concretely, the agent D and E have reached the same goal each other. This situation seems to be close, but that is incorrect. That is because the agents C and D have a conflict even if the agent D can change the route. 
In addition, DRL makes the agents explore almost all states and actions, while PS cannot make that. Therefore, PS cannot change the agents' reaching goals easily, that is, PS is hard to solve the conflict in the figure \ref{fig:qtables45}.
This suggests that the cooperation is very difficult along to the number of the agents and DRL can achieve the valuable cooperation.
\section{Conclusion} \label{sec:conclusion}
This paper proposed directionality reinforcement learning (DRL) as the complete decentralized multi-agent reinforcement learning method which can achieve cooperation based on each agent's learning: no communication and no observation. 
Concretely, DRL extends PMRL to limit the agents' reaching goals  by checking the mean of the acquired rewards.
That operation means adding the direction ``agents have to learn to reach the farthest goal among reachable ones" to learning agents to operate the agents cooperatively. 
Furthermore, to investigate the effectiveness of the DRL, this paper compare Q-learning agent with DRL with Profit sharing for multi-agent system as the previous method in maze problems. Experimental results derive that (1) DRL performs better than the previous method in terms of the spending time, (2) the direction makes agents learn yielding action for others, and (3) DRL suggests achieving multi-agent learning with few costs for any number of agents.

In this paper, DRL makes agents do two operations at one time, reaching the farthest goal and avoiding the non-first reaching goal. However, DRL has to prioritize the avoiding non-first reaching goal operation when the large number of the agents exist because of the results for five agents. We are going to establish the appropriate timing to do two operations in the future.
\begin{acks}
This research is supported by JSPS KAKENHI Grant Number JP17J08724.
\end{acks}

%%%%%%%%%%%%%%%%%%%%%%%%%%%%%%%%%%%%%%%%%%%%%%%%%%%%%%%%%%%%%%%%%%%%%%%%%%%%%%%%%%%%%%%%%%%%%%%%%%%%%%%%%
%% bibliography: see CFP for number of permitted pages

\bibliographystyle{ACM-Reference-Format}  % do not change this line!
\bibliography{aamas20}  % put name of your .bib file here

\end{document}